\documentclass[,final]{aipproc}
\layoutstyle{6x9}

\newcommand{\dsp}{\displaystyle}

\newcommand{\beq}{\begin{equation}}
\newcommand{\eeq}{\end{equation}}
\newcommand{\ba}{\begin{array}}
\newcommand{\ea}{\end{array}}
\newcommand{\bea}{\begin{eqnarray}}
\newcommand{\eea}{\end{eqnarray}}
\newcommand{\bal}{\begin{align}}  
\newcommand{\eal}{\end{align}}
\newcommand{\bi}{\begin{itemize}}  
\newcommand{\ei}{\end{itemize}}
\newcommand{\ben}{\begin{enumerate}}  
\newcommand{\een}{\end{enumerate}}

\newcommand\hide[1]{}

\newcommand{\tr}{\mbox{tr}}

\newcommand{\ds}[1]{
  \setbox0=\hbox{\ensuremath{#1}}
  \hbox to\wd0{\hbox to0pt{\hbox to\wd0{\hss/\hss}\hss}\box0}}

\begin{document}

\begin{flushright}
TKYNT-10-13, INT-PUB-10-044
\end{flushright}
\vspace*{-5ex}

\title{BEC-BCS crossover driven by the axial anomaly in the NJL model}


\author{Hiroaki Abuki}{
  address={Department of Physics, Tokyo University of Science,
  Tokyo 162-8601, Japan%
  },
}
\author{Gordon Baym}{
  address={Department of Physics, University of Illinois, 1110 W. Green St.,
Urbana, Illinois 61801, USA%
  }
}
\author{Tetsuo Hatsuda}{
  address={Department of Physics, The University of Tokyo, Tokyo
  113-0033, Japan%
  }
}
\author{Naoki Yamamoto}{
  address={Institute for Nuclear Theory, University of Washington,
  Seattle, WA 98195-1550, USA%
  }
}

\begin{abstract}
We study the QCD phase structure in the three-flavor Nambu--Jona-Lasinio 
model, incorporating the chiral-diquark interplay due to the axial
 anomaly.
We demonstrate that for a certain range of model parameters, 
the low temperature critical point predicted by a Ginzburg-Landau
 analysis
appears in the phase diagram. 
In addition, we show that the axial anomaly presents a new scenario for
 a possible BEC-BCS crossover in the color-flavor locked phase of QCD.
\end{abstract}

\maketitle

\section{Introduction}
The phases of QCD at finite temperature $T$ and quark chemical potential
$\mu$ 
are being actively studied.
In particular, at low $T$ and high $\mu$ a color superconducting (CSC)
phase \cite{reviews,Fukushima:2010bq} characterized by a diquark
condensate $\langle qq\rangle$ 
is expected to appear owing to the attractive interaction between
quarks, provided either by one-gluon exchange or by instantons. 
On the other hand, when the system at finite $\mu$ is heated,
a transition to a quark-gluon plasma (QGP) takes place at a
(pseudo-)critical temperature.

Our understanding of the phase structure based on the lattice simulations is
still immature due to the severe sign problem at finite $\mu$. Therefore
the analyses so far have relied mainly on specific models of QCD, such as
the Nambu--Jona-Lasinio model \cite{Hatsuda:1994pi,Buballa:2003qv}, the
Polyakov--Nambu--Jona-Lasinio (PNJL) model \cite{Fukushima:2003fw}, and
etc. Some of these model indicated the possible existence of a critical
point located at high $T$ \cite{Asakawa:1989bq}.

The interesting possibility of a second critical point at
rather low temperature in the color-flavor locked (CFL) phase was
recently predicted on the basis of general Ginzburg-Landau (GL) analysis
\cite{Hatsuda:2006ps}. Moreover, this critical point has proven to make
a quark-hadron continuity possible
\cite{Schafer:1998ef,Yamamoto:2007ah,Yamamoto:2008zw}.
In two-flavor QCD, similar critical points have been found in the
NJL model \cite{Kitazawa:2002bc},
although their origin is different from axial anomaly. 

This work is aimed at locating this critical point in 
the $(\mu,T)$-phase diagram using the phenomenological NJL model
\cite{Abuki:2010jq}.  
Starting with the three-flavor NJL model incorporating the axial anomaly
induced chiral-diquark interplay, we study the location of the 
new critical point and its dependence on the strength of the anomaly. 
We also observe that the axial anomaly triggers a crossover between a
Bose-Einstein condensed state (BEC) of diquark pairing and
Bardeen-Cooper-Schrieffer (BCS) diquark pairing \cite{Abuki:2001be,
Nishida:2005ds, He:2007kd,Kitazawa:2007zs,Baym:2008me} in the CFL phase. 

\section{Incorporating the axial anomaly in NJL model}
\label{sec:njl}
The Lagrangian of the NJL model with three-flavors consists of three
terms:
\beq
\label{eq:njl}
{\cal L}=
\bar q (i\gamma_{\mu}\partial^{\mu}-m_q  + \mu \gamma_0) q 
+ {\cal L}^{(4)} + {\cal L}^{(6)},
\eeq
where $q=({\rm u,d,s})$ is the flavor triplet quark field, 
$m_q$ 
is a flavor symmetric quark mass ($m_u=m_d=m_s$), and $\mu$ is the
chemical potential for conserved quark number.
${\cal L}^{(4)}$ and ${\cal L}^{(6)}$ are the 
four-fermion and six-fermion interactions, respectively.
As usual we set ${\cal L}^{(4)}={\cal L}^{(4)}_\chi+{\cal L}^{(4)}_d$
with the standard choice \cite{Hatsuda:1994pi,Buballa:2003qv}
\beq
{\cal L}^{(4)}_{\chi}
=8 G {\rm tr} ({\phi^{\dag} \phi}), 
\quad {\cal L}^{(4)}_{d}%
=2H {\rm tr} [d_L^{\dag}d_L + d_R^{\dag}d_R],
\eeq
where $\phi_{ij} \equiv (\bar{q}_R)^j_a (q_L)^i_a$,
$(d_L)_{ai} \equiv \epsilon_{abc} \epsilon_{ijk} ({q}_L)^j_b C (q_L)^k_c$,
and $(d_R)_{ai} \equiv \epsilon_{abc} \epsilon_{ijk} ({q}_R)^j_b C (q_R)^k_c$,
with $a, b, c$ and $i, j, k$  the color and flavor indices,
and $C$ the charge conjugation operator. 
The flavor $\mathrm{U}(3)$ generators $\tau_a$ ($a=0, \cdots, 8$) are
normalized so that ${\rm tr}[\tau_a \tau_b]=2\delta_{ab}$, and $\tau_A$
and $\lambda_{A'}$  with $A,A'=2,5,7$ are antisymmetric generators of
flavor and $\mathrm{SU}(3)$ color, respectively. ${\cal L}^{(4)}$ is
invariant under $\mathrm{SU}(3)_L\times\mathrm{SU}(3)_R \times
\mathrm{U}(1)_A \times \mathrm{U}(1)_B$
symmetry.  
The interaction ${\cal L}^{(4)}_{\chi}$ produces attraction 
of $q\bar{q}$ pairs, leading to the formation of a chiral condensate.
Similarly ${\cal L}^{(4)}_{d}$ leads to attraction of $qq$ pairs in the  
color-anti-triplet and spin-parity $0^{\pm}$ channel,  
inducing a color-flavor locked (CFL) condensate \cite{reviews}.
We treat the two couplings $G$ and $H$ as independent parameters.

The six-fermion interaction in our model consists of two parts,
${\cal L}^{(6)}={\cal L}^{(6)}_{\chi}+{\cal L}^{(6)}_{\chi d}$.
${\cal L}^{(6)}_{\chi}$ 
is the standard Kobayashi-Maskawa-'t Hooft (KMT) interaction
\cite{Kobayashi:1970ji}, 
\beq
\label{eq:kmt}
{\cal L}^{(6)}_{\chi}=- 8 K 
\left(\det \phi +{\rm h.c.}\right).
\eeq
This interaction is not invariant under $\mathrm{U}(1)_A$ symmetry,
which accounts for the axial anomaly in QCD due to instantons.
Consequently the mass of the $\eta'$ meson becomes larger than that of
the other pseudoscalar octet Nambu-Goldstone (NG) bosons ($\pi, \eta,
K$) 
for positive value of $K$.
On the other hand, the term (\ref{eq:kmt}) makes the chiral phase
transition first-order as a function of $T$ at $\mu=0$ for the massless
three-flavor limit \cite{Pisarski:1983ms}.

As shown in \cite{Hatsuda:2006ps}, the instanton can induce a coupling
between the chiral and diquark condensates through a new six-fermion
term:
\beq
\label{eq:K'}
{\cal L}^{(6)}_{\chi d} 
= K' \left(\tr[(d_R^{\dag} d_L) \phi] +{\rm h.c.} \right).
\eeq 
It is this term that is responsible for the aforementioned low
temperature critical point.
We assume $K'>0$, so that $qq$ pairs in the positive parity channel, 
$\langle d_L \rangle =-\langle d_R \rangle$, 
are energetically favored. We keep $K$ and $K'$ as independent
parameters \cite{Abuki:2010jq}.

The condensates favored by the interaction ${\cal L}^{(4)}+{\cal
L}^{(6)}$ 
are the flavor-symmetric chiral and diquark condensates in the
spin-parity $0^{+}$ channel, defined by
\beq
\langle \phi_{ij} \rangle
=(\chi/2) \delta_{ij}, \quad \langle d_{Lai}
 \rangle = -\langle d_{Rai} \rangle = (s/2) \delta_{ai}.
\eeq
Here the condensate order parameters are $\chi$ and $s$.

It is straightforward to derive the thermodynamic potential at the
mean-field level \cite{Abuki:2010jq}
\beq
\label{eq:thermo-pot}
\ba{rcl}
\Omega(\chi,s;\mu,T)&=&\dsp U(\chi,s)-\int_{|p|\le\Lambda}%
 \frac{d^3p}{(2\pi)^3}\sum_{\pm}
\left[ 
8 \omega_8^{\pm} + \omega_1^{\pm}\right]\\[2ex]
&&\dsp -2T\int\frac{d^3p}{(2\pi)^3}\sum_{\pm}
\left[8 \ln(1+e^{-\omega_8^{\pm}/T})+\ln(1+e^{-\omega_1^{\pm}/T})\right],
\ea
\eeq
where $\Lambda$ is a momentum cutoff to regulate the vacuum energy,
\beq
U(\chi,s) = 6G \chi^2 + 3H |s|^2 - 4K \chi^3 
- \frac{3}{2} K' |s|^2 \chi,
\label{eq:V}
\eeq
is a constant term which is needed to cancel double counting of the
interactions, and
\beq
\label{eq:dispersion}
\omega_8^{\pm}=\sqrt{(\sqrt{M^2+p^2} \pm \mu)^2+ (2\Delta)^2}, \quad
\omega_1^{\pm}=\sqrt{(\sqrt{M^2+p^2} \pm \mu)^2+ \Delta^2}
\eeq
are the dispersion relations for the quasi-quarks in the octet and singlet
representations, with $M$ and $\Delta$ the dynamical Dirac and
Majorana masses, defined as
\beq
M = m_q-4 \left(G-\frac{1}{2}K \chi \right) \chi %
 + \frac{1}{4}{K'}|s|^2,\quad
\Delta = -2 \left(H - \frac{1}{4} K' \chi \right)|s|.
\eeq
These equations imply that $\chi<0$ is energetically favored for
non-zero $m_q$, while $s$ is generally complex; the thermodynamic
potential is a function of $|s|^2$.

\section{Phase structure and Discussion}
\label{sec:phase}
The phase structures can be determined numerically by looking for the
values of $\chi$ and $s$ that minimize the thermodynamic potential 
in Eq.~(\ref{eq:thermo-pot}). 
We follow the parameter choice of \cite{Buballa:2003qv}.
We show in Table \ref{tab} two sets of parameters we adopt, Set I and
Set II respectively.
We vary the strength of the  chiral-diquark coupling (the $K'$ term) by hand.
We work in the flavor $\mathrm{SU}(3)$ limit, assuming 
$m_u=m_d=m_s\equiv m_q$ 
for simplicity.
\begin{table}[t]
\begin{tabular}{|c|c|c|c|c|c|c|}
\hline 
& $m_q$ [MeV] &  $G\Lambda^2$  &  $H\Lambda^2$ 
&  $K\Lambda^5$  & $M$ [MeV] & $\chi^{1/3}$ [MeV] \\ \hline \hline
 Set I  & 0 & 1.926 & 1.74 & 12.36 & 355.2 & $-$240.4 \\ \hline
 Set II  & 5.5 & 1.918 & 1.74 & 12.36 & 367.6 & $-$241.9  \\  \hline
\end{tabular}
\caption{Two sets of parameters in the present three-flavor NJL model: 
The momentum cutoff is fixed at $\Lambda=602.3$ MeV \cite{Buballa:2003qv}.
The dynamical quark mass $M$ and the chiral condensate $\chi$ at vacuum 
are also given. }
\label{tab}
\end{table}

\begin{figure}[t]
\centerline{
\includegraphics[width=0.48\textwidth]{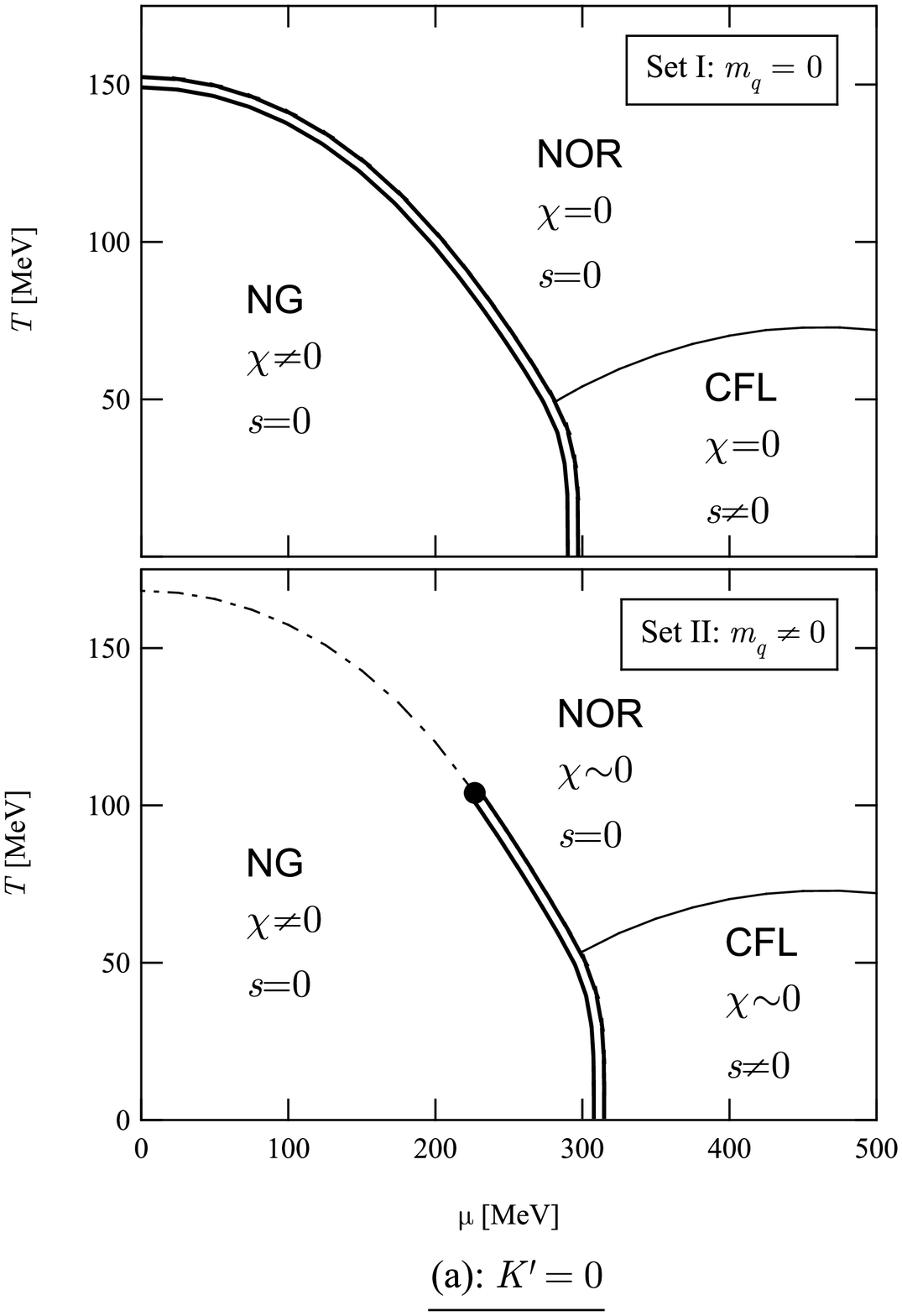}
\includegraphics[width=0.48\textwidth]{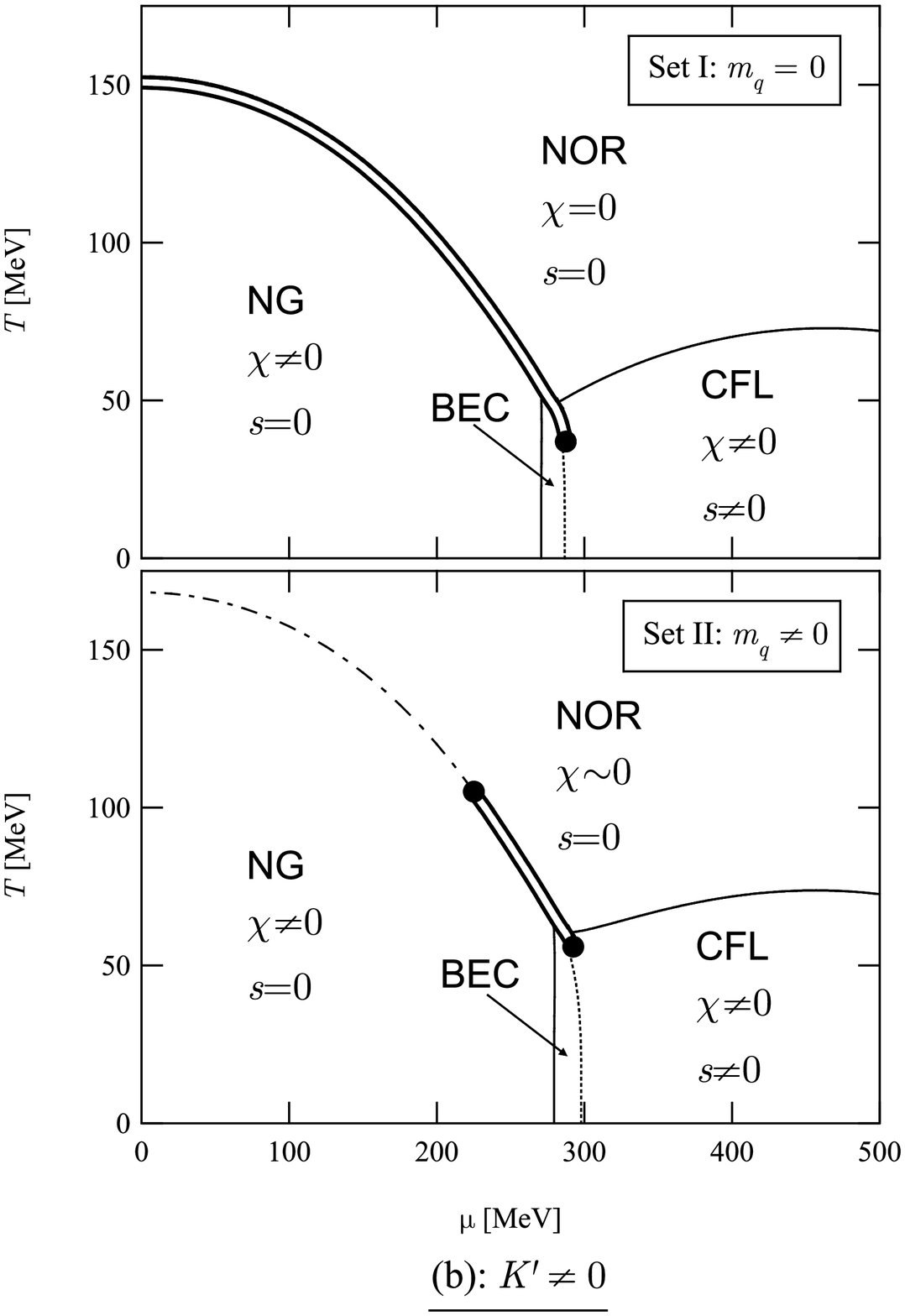}
}

\caption{%
The phase structure in the ($\mu,T$)-plane in
the three-flavor NJL model without (a) and with (b) the $K'$ term.
The Upper and lower panels present the results in the massless case I
 and the massive case II respectively.
Phase boundaries with a second-order transition are denoted 
by a single line and a first-order transition by a double line. 
The dashed-dot line at high T in case II shows the chiral
 crossover line, while the dotted line in (b) denotes the BEC-BCS
 crossover. See \cite{Abuki:2010jq} for further details.
}
\label{fig:njl0}
\end{figure}

We show in Fig.~\ref{fig:njl0} the phase structures for Set I (massless
case) in the upper panel, and those for Set II (massive case) in the
lower panel. Panels (a) and (b) show the results without and 
with the $K'$-term; in (b) we have taken $K'=4.2K_0$ with
$K_0=12.36/\Lambda^5$ 
as a representative value.
The phase diagrams contain a CFL phase with $s\ne 0$ 
with $\mathrm{U}(1)$ baryon number broken, and other two phases both
characterized with $s=0$, a Nambu-Goldstone (NG) phase with $\chi\ne 0$
and a normal (NOR) phase with either $\chi=0$ (in case $m_q=0$) 
or $\chi\sim 0$ (in case $m_q\ne0$).
From the two figures in the panel (a), we see that the current
quark mass leads to the critical point on the high temperature side 
of the first-order line of chiral phase transition
\cite{Asakawa:1989bq}.
The critical point moves downwards with increasing quark mass $m_q$
since it acts as an external symmetry breaking source of chiral symmetry
breaking and thus smears the strength of the phase transition.

The effect of nonvanishing $K'$ can be seen by comparing (a) and (b).
We indeed see that the low temperature critical point shows up at the
other end of the line of the first order chiral phase transition.
This is , as discussed in \cite{Hatsuda:2006ps}, 
because the $K'$-term acts as an external field for $\chi$,
which turns the first-order chiral phase transition into a crossover
in the CFL phase where $s\ne 0$. 
Note that the CFL phase in the panel (b) accompanies a nonzero chiral
condensate $\chi\ne0$ induced by the anomaly mixing 
term ${\cal L}^{(6)}_{\chi d}$.

The axial anomaly,  for sufficiently large chiral-diquark coupling $K'$,
not only triggers the low $T$  critical point, but also drives a BEC-BCS
crossover in the CFL phase, as discussed in \cite{Baym:2008me}.  
Within an NJL-type model such a BEC regime appears for sufficiently
large pairing attraction, $H$, in the $qq$-channel
\cite{Nishida:2005ds,He:2007kd,Kitazawa:2007zs}. 
The novel feature here is that the axial anomaly helps to realize the
BEC regime through its contribution to the effective $qq$ coupling. 
This can be easily seen by extracting from Eq.~(\ref{eq:K'}) 
the dominant zero mode ($\phi_{ij}\sim\delta_{ij}\chi/2$) contribution
to the quark-quark interaction
\beq
\ba{rcl}
{\cal L}^{(6)}_{\chi d}&\sim&\frac{1}{4}K'|\chi|%
 \tr\left[(d_R-d_L)^\dagger(d_R-d_L)\right]
 -\frac{1}{4}K'|\chi|\tr\left[(d_R+d_L)^\dagger(d_R+d_L)\right].
\ea
\label{heff}
\eeq 
The first term increases the effective attraction between quarks
in the $0^+$ channel, while the second term is repulsive and suppresses
the $0^-$ pairing. Thus when the chiral condensate is nonvanishing, as
in the NG phase, the axial anomaly helps the formation of
a diquark BEC condensate.

In fact it is possible to show that at sufficiently large $K'$
there are nine diquark bound states with mass $M_D(\mu,T)\le 2M(\mu,T)$
where $M(\mu,T)$ is the dynamical Dirac mass of quarks at $\mu$ and $T$. 
Each diquark complex scalar has quark number $\pm 2$ so that it feels
chemical potential $2\mu$. Thus when $2\mu$ hits $M_D(\mu,T)$ from below
a BEC condensate must start to form. Then the condition for the
onset of a BEC approaching from the NG phase 
(NG-BEC boundary) is given by the condition 
\cite{nozieres,Nishida:2005ds}
\begin{eqnarray}
2 \mu = M_D(\mu,T).
\end{eqnarray}  
\begin{figure}[t]
\centerline{
\includegraphics[width=0.45\textwidth]{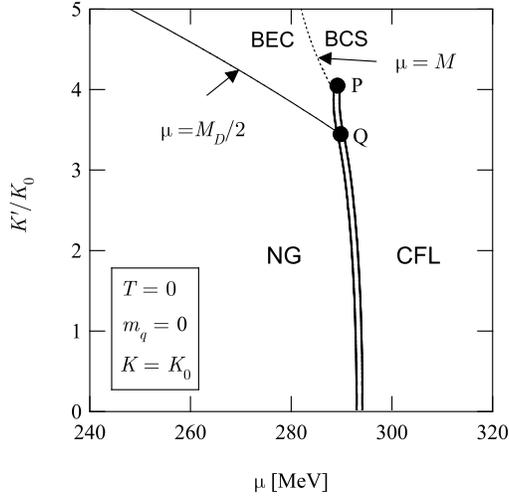} 
}
\caption{%
The phase diagram in the $(\mu,K')$-plane at $T=0$ for massless quarks,
 with the NG and CFL phases. A dotted line separates the CFL phase into
 BCS- and BEC-like domains. The critical point and the
 critical end point are denoted by P and Q, respectively.}
\label{fig:muKprime}
\end{figure}
In order to see how the BEC domain in the CFL phase grows as a function
of $K'$, we show in Fig.~\ref{fig:muKprime} the phase diagram in the
 ($\mu,K'$)-plane for massless quarks. 
The first-order line separating the CFL and NG phases for small $K'$
eventually terminates at the critical point P. 
On the other hand, for $K'$ sufficiently large, a BEC regime of bound
diquarks appears across a second-order phase transition at a critical
chemical potential $\mu=M_D(\mu,0)/2$ shown by solid line; the phase
boundary meets the first-order line at the critical end point Q.
A novel first-order transition from the BEC to BCS regimes appears
between P and Q, with discontinuous changes of both the chiral and
diquark condensates.

In conclusion, the axial anomaly, by driving a coupling between the
chiral and diquark condensates, plays an important role in the many 
body physics of QCD, making the phase diagram extremely rich.
For one, we demonstrated that it can indeed produce the low
temperature critical point between the hadronic phase and the color
superconducting phase predicted by the previous GL analysis
\cite{Hatsuda:2006ps}.
In addition, we have shown that the coupling helps
the formation of a BEC of diquarks via increasing effective 
quark-quark attraction in the $0^+$ channel. 
As a result, a BEC-BCS crossover or even the first
order BEC-BCS transition can be realized in the CFL phase.

Finally we note that very recently the extension of our analysis
incorporating the effect of heavy strange quark mass was reported
 \cite{Basler:2010xy}.
It still remains an important task to extend the analyses imposing the
charge neutrality and $\beta$-equilibrium conditions. 

\vspace*{1ex}
\noindent
The numerical calculations were carried out on Altix3700 at YITP in
 Kyoto University.

\bibliographystyle{aipproc}   


\begin{thebibliography}{99}

\bibitem{reviews}
  K.~Rajagopal and F.~Wilczek,
  arXiv:hep-ph/0011333;
  D.~H.~Rischke,
  Prog.\ Part.\ Nucl.\ Phys.\  {\bf 52}, 197 (2004);
  M.~G.~Alford, A.~Schmitt, K.~Rajagopal and T.~Schafer,
  Rev.\ Mod.\ Phys.\  {\bf 80}, 1455 (2008).

\bibitem{Fukushima:2010bq}
  K.~Fukushima and T.~Hatsuda,
  arXiv:1005.4814 [hep-ph].

\bibitem{Hatsuda:1994pi}
  T.~Hatsuda and T.~Kunihiro,
  Phys.\ Rept.\  {\bf 247}, 221 (1994).

\bibitem{Buballa:2003qv}
  M.~Buballa,
  Phys.\ Rept.\  {\bf 407}, 205 (2005).

\bibitem{Fukushima:2003fw}
  K.~Fukushima,
  Phys.\ Lett.\  B {\bf 591}, 277 (2004);
  C.~Ratti, M.~A.~Thaler and W.~Weise,
  Phys.\ Rev.\  D {\bf 73}, 014019 (2006);
  S.~Roessner, C.~Ratti and W.~Weise,
  Phys.\ Rev.\  D {\bf 75}, 034007 (2007);
  H.~Abuki, M.~Ciminale, R.~Gatto, G.~Nardulli and M.~Ruggieri,
  Phys.\ Rev.\  D {\bf 77}, 074018 (2008)
  [arXiv:0802.2396 [hep-ph]];
  H.~Abuki, R.~Anglani, R.~Gatto, G.~Nardulli and M.~Ruggieri,
  Phys.\ Rev.\  D {\bf 78}, 034034 (2008)
  [arXiv:0805.1509 [hep-ph]].

\bibitem{Asakawa:1989bq}
  M.~Asakawa and K.~Yazaki,
  Nucl.\ Phys.\  A {\bf 504} (1989) 668;
  A.~Barducci, R.~Casalbuoni, S.~De Curtis, R.~Gatto and G.~Pettini,
  Phys.\ Lett.\  B {\bf 231}, 463 (1989).

\bibitem{Hatsuda:2006ps}
  T.~Hatsuda, M.~Tachibana, N.~Yamamoto and G.~Baym,
  Phys.\ Rev.\ Lett.\  {\bf 97}, 122001 (2006).

\bibitem{Schafer:1998ef}
  T.~Schafer and F.~Wilczek,
  Phys.\ Rev.\ Lett.\  {\bf 82}, 3956 (1999)
  [arXiv:hep-ph/9811473].

\bibitem{Yamamoto:2007ah}
  N.~Yamamoto, M.~Tachibana, T.~Hatsuda and G.~Baym,
  Phys.\ Rev.\  D {\bf 76}, 074001 (2007);
  T.~Hatsuda, M.~Tachibana and N.~Yamamoto,
  Phys.\ Rev.\  D {\bf 78}, 011501 (2008).

\bibitem{Yamamoto:2008zw}
  N.~Yamamoto,
  JHEP {\bf 0812}, 060 (2008).

\bibitem{Kitazawa:2002bc}
  M.~Kitazawa, T.~Koide, T.~Kunihiro and Y.~Nemoto,
  Prog.\ Theor.\ Phys.\  {\bf 108}, 929 (2002).

\bibitem{Abuki:2010jq}
  H.~Abuki, G.~Baym, T.~Hatsuda and N.~Yamamoto,
  Phys.\ Rev.\  D {\bf 81}, 125010 (2010)
  [arXiv:1003.0408 [hep-ph]].

\bibitem{Abuki:2001be}
  H.~Abuki, T.~Hatsuda and K.~Itakura,
  Phys.\ Rev.\  D {\bf 65}, 074014 (2002)
  [arXiv:hep-ph/0109013].

\bibitem{Nishida:2005ds}
  Y.~Nishida and H.~Abuki,
  Phys.\ Rev.\  D {\bf 72}, 096004 (2005)
  [arXiv:hep-ph/0504083];
  H.~Abuki,
  Nucl.\ Phys.\  A {\bf 791}, 117 (2007)
  [arXiv:hep-ph/0605081].

\bibitem{He:2007kd}
  L.~He and P.~Zhuang,
  Phys.\ Rev.\  D {\bf 75}, 096003 (2007);
  L.~He and P.~Zhuang,
  Phys.\ Rev.\  D {\bf 76}, 056003 (2007);
  L.~He,
  arXiv:1007.1920 [hep-ph];
  J.~c.~Wang, Q.~Wang and D.~H.~Rischke,
  arXiv:1008.4029 [nucl-th].


\bibitem{Kitazawa:2007zs}
  M.~Kitazawa, D.~H.~Rischke and I.~A.~Shovkovy,
  Phys.\ Lett.\  B {\bf 663}, 228 (2008).

\bibitem{Baym:2008me}
  G.~Baym, T.~Hatsuda, M.~Tachibana and N.~Yamamoto,
  J.\ Phys.\ G {\bf 35}, 104021 (2008).


\bibitem{Kobayashi:1970ji}
  M.~Kobayashi and T.~Maskawa,
  Prog.\ Theor.\ Phys.\  {\bf 44} (1970) 1422;
  G.~'t Hooft,
  Phys.\ Rev.\ Lett.\  {\bf 37}, 8 (1976);
  Phys.\ Rev.\  D {\bf 14}, 3432 (1976)
  [Erratum-ibid.\  D {\bf 18}, 2199 (1978)].

\bibitem{Pisarski:1983ms}
  R.~D.~Pisarski and F.~Wilczek,
  Phys.\ Rev.\  D {\bf 29}, 338 (1984).

\bibitem{nozieres}
  P.~Nozi{\`e}res and S.~Schmitt-Rink, 
  J.\ Low.\ Temp.\ Phys.\ {\bf 59} 195 (1985).

\bibitem{Basler:2010xy}
  H.~Basler and M.~Buballa,
  arXiv:1007.5198 [hep-ph].

\end{thebibliography}

\end{document}